\documentclass[letterpaper,11pt]{article}

\usepackage{scrextend}
\usepackage{amsmath,amssymb,epsfig,bbm}
\usepackage[active]{srcltx}
\usepackage[all]{xypic}
\usepackage{MnSymbol}
\usepackage{slashed}
\usepackage{amsthm}

\usepackage{float}

\usepackage{color}

\usepackage{dsfont}

\definecolor{co}{cmyk}{0,0.7,0.3,0}
\definecolor{darkgreen}{cmyk}{1,0,1,.2}
\definecolor{m}{rgb}{1,0.1,1}

\newcommand{\be}{\begin{equation}}
\newcommand{\ba}{\begin{eqnarray}}
\newcommand{\ea}{\end{eqnarray}}
\newcommand{\nn}{\nonumber}

\def\a{\alpha}
\def\b{\beta}

\def\d{\delta}
\def\e{\epsilon}

\def\m{\mu}
\def\n{\nu}
\def\oo{\omega}
\def\p{\pi}

\def\s{\sigma}

\def\x{\xi}

\def\G{\Gamma}

\def\OO{\Omega}
\def\P{\Pi}

\def\ca{{\cal A}}
\def\cb{{\cal B}}

\def\ch{{\cal H}}

\def\cn{{\cal N}}

\def\cs{{\cal S}}

\makeatletter
\newcommand{\eqnum}{\refstepcounter{equation}\textup{\tagform@{\theequation}}}
\makeatother

\newcommand{\pa}{\partial}

\newtheorem{thm}{Theorem}[subsection]

\newtheorem*{definition*}{Definition}

\fontfamily{yfrak}

\begin{document}

\vskip 25mm

\begin{center}

{\large\bfseries  


Nonperturbative Quantum Field Theory\\ and Noncommutative Geometry

}

\vskip 6ex

Johannes \textsc{Aastrup}$^{a}$\footnote{email: \texttt{aastrup@math.uni-hannover.de}} \&
Jesper M\o ller \textsc{Grimstrup}$^{b}$\footnote{email: \texttt{jesper.grimstrup@gmail.com}}\\ 
\vskip 3ex

$^{a}\,$\textit{Mathematisches Institut, Universit\"at Hannover, \\ Welfengarten 1, 
D-30167 Hannover, Germany.}
\\[3ex]
$^{b}\,$\textit{QHT Gruppen, Copenhagen, Denmark.}
\\[3ex]

{\footnotesize\it This work is financially supported by Ilyas Khan, \\\vspace{-0.1cm}St EdmundÕs College, Cambridge, United Kingdom.}

\end{center}

\vskip 3ex

\begin{abstract}

A general framework of non-perturbative quantum field theory on a curved background is presented. A quantum field theory is in this setting characterised by an embedding of a space of field configurations into a Hilbert space over $\mathbb{R}^\infty$. This embedding, which is only local up to a scale that we interpret as the Planck scale, coincides in the local and flat limit with the plane wave expansion known from canonical quantisation. We identify a universal Bott-Dirac operator acting in the Hilbert space over $\mathbb{R}^\infty$ and show that it gives rise to the free Hamiltonian both in the case of a scalar field theory and in the case of a Yang-Mills theory. These theories come with a canonical fermionic sector for which the Bott-Dirac operator also provides the Hamiltonian. We prove that these quantum field theories exist non-perturbatively for an interacting real scalar theory and for a general Yang-Mills theory, both with or without the fermionic sectors, and show that the free theories are given by semi-finite spectral triples over the respective configuration spaces. Finally, we propose a class of quantum field theories whose interactions are generated by inner fluctuations of the Bott-Dirac operator.

\end{abstract}

\newpage
\tableofcontents

\section{Introduction}
\setcounter{footnote}{0}


In quantum mechanics on for example the real line $\mathbb{R}$ one usually considers $L^2(\mathbb{R})$ together with the operators $x$ and $i\frac{d}{dx}$ satisfying the Heisenberg relation
\begin{equation}
[i \frac{d}{dx},x] = i.
\label{no1}
\end{equation}
Alternatively one can replace the operator $x$ with $C_c^\infty(\mathbb{R})$, the space of smoothly supported compact functions, and the operator $i \frac{d}{dx}$ with translations in $\mathbb{R}$, i.e. by the operators
$$
U_a f(x) = f(x-a),\quad a\in\mathbb{R}, f\in L^2(\mathbb{R}),
$$
which satisfy the relation
\begin{equation}
U_a f U_a^*  (x) = f(x+a).
\label{no2}
\end{equation}
These two formulations of quantum mechanics -- let us call them the {\it infinitesimal} and the {\it integrated} formulations, respectively -- are of course equivalent but only as long as the number of degrees of freedom remains finite: as we shall show in this paper, the representation theory of the integrated formulation is, when we consider fields, richer than that of the infinitesimal formulation. This  realisation is the starting point of the framework of non-perturbative quantum field theory, that we present in this paper.


Thus, instead of searching for algebraic representations of the canonical commutation relations of field operators we shall instead identify algebras of bounded functions over spaces of field configurations together with translation operators hereon. Such algebras turn out to have nontrivial Hilbert space representations, which are {\it not} accessible via the infinitesimal formulation and which allow for a construction of interacting quantum field theories on curved manifolds. 

The reason why an approach to non-perturbative quantum field theory based on the integrated formulation -- i.e. on algebras of functions on configuration spaces together with translation operators -- is advantageous is that it permits non-local representations, where the canonical commutation relations are only realised up to a correction at a scale, which we interpret as the Planck scale. Thus, these representations depend on a scale, which prevents arbitrary localisation, and therefore they do {\it not} include the operator valued distributions known from perturbative quantum field theory.

The key step in finding these Hilbert space representations is to construct a measure, where fast oscillating field configurations have smaller probabilities than the slow oscillating ones. Concretely, we do this by expanding the fields in terms of eigenvalues of a Laplace operator and then constructing a Gaussian measure on the space of the coefficients of this expansion weighted with the eigenvalues of the Laplace operator. The result is that transitions between field configurations depend on a Sobolev norm,  which is inherently non-local.\\

In this paper we present both a general method of constructing non-perturbative quantum field theories as well as two concrete examples. First we consider the case of a real scalar field and second we revisit the case of a gauge field, which was first developed in \cite{Aastrup:2017vrm,Aastrup:2017xde}. In both cases we find that the operator expansion in terms of weighted eigenvectors of a Laplace operator coincide in a local and flat limit with the plane wave expansion of a canonically quantised field. This means that the general framework, that we present, includes perturbative quantum field theory as a limiting case.

At the heart of our construction is a Hilbert space over $\mathbb{R}^\infty$ and a canonical Bott-Dirac operator, due to Higson and Kasparov \cite{Higson}, that acts on functions on $\mathbb{R}^\infty$. The weighted expansion of fields in eigenfunctions of a Laplace operator is an embedding from the space of field configurations into $\mathbb{R}^\infty$, the space of coefficients in this expansion. The square of the Bott-Dirac operator gives the Hamilton operator of an infinite-dimensional harmonic oscillator, which in turn gives -- both in the case of a real scalar field and in case of a gauge field -- the free Hamiltonian of the corresponding quantum field theory in the flat and local limit. 
Furthermore, the construction of the Bott-Dirac operator naturally introduces also a fermionic sector to the theories we study, where the square of the Bott-Dirac operator gives the Hamilton operator for the fermions as well.

We also consider the case of interacting quantum field theories. Here the most important feature is that these theories continue to exist also when interactions are turned on (irrespectively on whether the fermions are included or not). 


The Bott-Dirac operator is a canonical geometrical structure acting in a universal $L^2$ space over various spaces of field configurations, where it forms a semi-finite spectral triple. The representations, that we have found, involve, however, also an $L^2$ space over the three-dimensional manifold $M$ and it is therefore natural to consider a also Dirac-type operator, that consist of two parts: the Bott-Dirac operator plus a spatial Dirac operator acting on spinors on the manifold. Such an operator interacts also with the various representation of the field operators and forms again a spectral triple except that commutators with the algebra are no longer bounded.

One particularly interesting construction emerges when one considers the algebra of field operators for a gauge theory -- this is what we call the holonomy-diffeomorphism algebra, denoted $\mathbf{HD}(M)$, since it is generated by holonomies along flows of vector-fields \cite{AGnew}  -- and let it act on spinors in $L^2(M,S)$. This is what we have previously called {quantum holonomy theory} \cite{Aastrup:2017xde,Aastrup:2015gba}. 
Perhaps the most interesting feature of this model is that it has a possible connection to the standard model of particle physics via its formulation in terms of noncommutative geometry due to Chamseddine and Connes  \cite{Connes:1996gi,Chamseddine:2007hz}. The point is that the $\mathbf{HD}(M)$ algebra produces a so-called almost commutative algebra in a semi-classical limit \cite{Aastrup:2012vq}, which is the type of algebra that Chamseddine and Connes have identified as a basic geometrical input in their work on the standard model coupled to general relativity.\\

The idea to construct a Dirac-type operator over a space of field configurations and to let it interact with an algebra of field operators was first proposed in \cite{Aastrup:2005yk} and later developed in \cite{Aastrup:2012jj}-\cite{Aastrup:2008wb}. The $\mathbf{HD}(M)$ algebra was first introduced in \cite{AGnew,Aastrup:2012vq} and further studied in \cite{Aastrup:2017vrm,Aastrup:2017xde,Aastrup:2015gba,Aastrup:2016caz,Aastrup:2016ytt,Aastrup:2014ppa}. After completing the papers  \cite{Aastrup:2017vrm,Aastrup:2017xde} we became aware of the work by Higson and Kasparov \cite{Higson}, where a similar Hilbert space construction was used and where the infinite-dimensional Bott-Dirac operator was first introduced -- the operator, which is the central object in the present paper.\\

We begin in the next section by introducing the infinite-dimensional Bott-Dirac operator acting in a Hilbert space over the projective limit $\mathbb{R}^\infty$. In section 3 we then outline the general construction of a non-perturbative quantum field theory based on the Bott-Dirac operator. Here the key step is the embedding of the space of field configurations into the space $\mathbb{R}^\infty$. With this we then move on to discuss the first example, in section 4, with a real scalar field. We construct a Hilbert space representation of an algebra of field operators and prove that it exist and is strongly continuous. We also show that our construction coincides with perturbative quantum field theory in a flat and local limit. Finally we show that the Bott-Dirac operator together with the algebra of field operators form a semi-finite spectral triple over the space of scalar field configurations. Next, in section 5, we turn to gauge theory, where we introduce the $\mathbf{HD}(M)$ and $\mathbf{QHD}(M)$ algebras and their representations. Again we show that our construction coincides with perturbative quantum field theory in a flat and local limit and that the Bott-Dirac operator together with the $\mathbf{HD}(M)$ algebra form a semi-finite spectral triple. In section 6 we show that the construction of the Bott-Dirac operator naturally introduces a fermionic sector and  that its square again gives the Hamiltonian. Furthermore, in section 7 we add a spatial Dirac operator to the Bott-Dirac operator and consider, in section 8, fluctuations of the Bott-Dirac operator by inner automorphisms and show how this generates a class of interacting quantum field theories. In section 9 we then argue that the non-local nature of the quantum field theories, that we find, puts into question one of the strongest arguments in favour of a quantum theory of gravity. We end the paper in section 10 with a discussion.

\section{An infinite-dimensional Bott-Dirac operator}

We begin with a basic geometrical construction, which in the next sections will play a key role in the formulation of various quantum field theories. The following formulation of an infinite-dimensional Bott-Dirac operator is due to Higson and Kasparov \cite{Higson}.\\

Let $\ch_n= L^2(\mathbb{R}^n)$, where the measure is given by the flat metric, and consider the embedding
$$
\varphi_n : \ch_n\rightarrow\ch_{n+1}
$$
given by 
\begin{equation}
\varphi_n(\eta)(x_1,x_2,\ldots x_{n+1}) = \eta(x_1,\ldots, x_n)  \left(\frac{s_{n+1}}{\tau_2\pi}\right)^{1/4}e^{- \frac{s_{n+1} x_{n+1}^2}{2\tau_2}},
\label{ref}
\end{equation}
where $\{s_n\}_{n\in\mathbb{N}}$ is a monotonously increasing sequence of parameters, which we for now leave unspecified\footnote{In \cite{Higson} these parameters were not included, i.e. $s_n=1\forall n$.  }. This gives us an inductive system of Hilbert spaces 
$$
\ch_1\stackrel{\varphi_1}{\longrightarrow}  \ch_2 \stackrel{\varphi_2}{\longrightarrow}  \ldots   \stackrel{\varphi_n}{\longrightarrow}    \ch_{n+1} \stackrel{\varphi_{n+1}}{\longrightarrow}\ldots
$$
and we define\footnote{The notation $L^2 (\mathbb{R}^\infty )$, which we are using here, is somewhat ambiguous. We are here only considering functions on  $\mathbb{R}^\infty $ with a specific tail behaviour, namely the one generated by (3). We have not included this  tail behaviour in the notation. See \cite{Aastrup:2017vrm} for further details.}
 $L^2(\mathbb{R}^\infty) $ as the Hilbert space direct limit
\begin{equation}
L^2(\mathbb{R}^\infty) = \lim_{\rightarrow} L^2(\mathbb{R}^n)
\end{equation}
taken over the embeddings $\{\varphi_n\}_{n\in\mathbb{N}}$ given in (\ref{ref}).
%
We are now going to define the Bott-Dirac operator on $ L^2(\mathbb{R}^n)\otimes \Lambda^*\mathbb{R}^n$. 
Denote by $\mbox{ext}(f)$ the operator of external multiplication with $f$ on $\Lambda^*\mathbb{R}^n$, where $f$ is a vector in $\mathbb{R}^n$, and denote by $\mbox{int}(f)$ its adjoint, i.e. the interior multiplication by $f$.
Denote by $\{v_i\}$ a set of orthonormal basis vectors on $\mathbb{R}^n$ and let $\bar{c}_i$ and $c_i$ be the Clifford multiplication operators given by
\begin{eqnarray}
{c}_i &=& \mbox{ext}(v_i) + \mbox{int}(v_i)
\nn\\
\bar{c}_i &=& \mbox{ext}(v_i) - \mbox{int}(v_i) 
\end{eqnarray}
that satisfy the relations 
\begin{eqnarray}
 \{c_i, \bar{c}_j\} = 0, \quad
 \{c_i, {c_j}\} = \d_{ij}, \quad
 \{\bar{c}_i, \bar{c}_j\} =- \d_{ij}.
\end{eqnarray}
We shall also use the notation:
\begin{equation}
\left.
\begin{array}{c}
\mathfrak{a}^\dagger_i:= \mbox{ext}(v_i)
\\
 \mathfrak{a}_i:=  \mbox{int}(v_i)
\end{array}
\right\}
\quad
\mbox{with}
\quad
\{\mathfrak{a}_i,\mathfrak{a}_j^\dagger\} = \d_{ij}.
\label{rolignu}
\end{equation}
The Bott-Dirac operator on $ L^2(\mathbb{R}^n)\otimes \Lambda^*\mathbb{R}^n$ is given by
$$
B_n = \sum_{i=1}^n\left( \tau_2 \bar{c}_i   \frac{\pa}{\pa x_i} +  s_i c_i  x_i\right),
$$
and the square of $B_n$ is
\begin{equation}
B_n^2 = \sum_{i=1}^n   \left(- \tau_2 \frac{\pa^2}{\pa x_i^2} + s_i^2 x_i^2\right) +  \tau_2 2\tilde{N} - \tau_2 \tilde{s} 
\label{udsigt}
\end{equation}
where $\tilde{N}$ is the operator 
$$
\tilde{N}= \sum_{i=1}^{n}s_i   \mathfrak{a}^\dagger_i \mathfrak{a}_i 
$$ 
which would, if we removed the factors $s_i$, be the operator that assigns the degree to a differential form in $\Lambda^*\mathbb{R}^n$, and where $\tilde{s}=\sum_{i=1}^n  s_i $.
With $B_n$ we can then construct the Bott-Dirac operator $B$ on $L^2(\mathbb{R}^\infty)\otimes \Lambda^*\mathbb{R}^\infty$ that coincides with $B_n$ on any finite subspace $L^2(\mathbb{R}^n)$. Here we mean by $\Lambda^*\mathbb{R}^\infty$ the inductive limit
$$
\Lambda^*\mathbb{R}^\infty= \lim_{\rightarrow} \Lambda^*\mathbb{R}^n.
$$
For details on the construction of $B$ we refer the reader to \cite{Higson}. 

We note that when we let the sequence $\{s_i \}_{i\in\mathds{N}}$ tend to infinity the eigenvalues of the Bott-Dirac operator will tend to infinity. This means that, modulo the infinite degeneracy coming from $\Lambda (\mathbb{R}^\infty )$, the eigen-spaces are finite dimensional. This means that modulo $\Lambda (\mathbb{R}^\infty )$ the Bott-Dirac operator $B$ has compact resolvent\footnote{Note that the sequence $\{s_i\}_{i\in\mathds{N}}$ was not included in the construction of the Bott-Dirac operator in \cite{Higson}. Instead the Bott-Dirac operator was there combined with a second operator with compact resolvent, the result being again a total operator with compact resolvent.}, i.e.
$
(B\pm i)^{-1}
$
is a compact operator.

We shall call the state
$$
\eta_{\mathrm{\bf gs}}  ({\bf x}) = \prod_i  \left(\frac{s_{i}}{\tau_2\pi}\right)^{1/4}e^{- \frac{s_{i} x_{i}^2}{2\tau_2}} \in L^2(\mathbb{R}^\infty)
$$
for the ground state. Note that $\eta_{\mathrm{\bf gs}} $ lies in the kernel of $B$, i.e.
$$
B ( \eta_{\mathrm{\bf gs}}  ) =0.
$$

If we define the creation and annihilation operators
\begin{eqnarray}
q_i = \sqrt{s_i} x_i +  \frac{\tau_2}{\sqrt{s_i}} \frac{\pa}{\pa x_i}
,\qquad
q_i\dagger = \sqrt{s_i} x_i -   \frac{\tau_2}{\sqrt{s_i}} \frac{\pa}{\pa x_i},
\end{eqnarray}
with
\begin{eqnarray}
x_i = \frac{1}{2\sqrt{s_i}} \left( q_i + q_i^\dagger \right)
,\qquad
\frac{\pa}{\pa x_i} = \frac{\sqrt{s_i}}{2\tau_2} \left( q_i - q_i^\dagger \right),
\end{eqnarray}
then we can rewrite $B$ as 
$$
B = \sum_i \sqrt{s_i}\left(  q^\dagger_i \mathfrak{a}_i  + q_i \mathfrak{a}_i^\dagger \right)
$$
and its square as
\begin{equation}
B^2 = \sum_i s_i\left(  q_i^\dagger q_i   + 2 \mathfrak{a}_i^\dagger  \mathfrak{a}_i      \right) .
\label{bingo}
\end{equation}
We then have the relations
\begin{eqnarray}
q_i =\frac{1}{\sqrt{s_i}} \{B , \mathfrak{a}_i\},&& \quad  q_i^\dagger = \frac{1}{\sqrt{s_i}}  \{B , \mathfrak{a}^\dagger\},
\nn\\
-2\mathfrak{a}_i = \frac{1}{\sqrt{s_i}}  [B, q_i], &&\quad2 \mathfrak{a}_i^\dagger = \frac{1}{\sqrt{s_i}}  [B, q_i^\dagger].
\end{eqnarray}
With these relations it is easy to see how the Bott-Dirac operator communicates between the CAR and the CCR algebras generated by $\{\mathfrak{a}_i^\dagger,\mathfrak{a}_i  \}$ and $(q^\dagger_i, q_i)$ respectively.

Let us finally introduce an additional piece of notation, that shall become useful later, namely
\begin{eqnarray}
 B^2\big\vert_{\mbox{\tiny b}} := \sum_i s_i q_i^\dagger q_i   
 ,\quad
 B^2\big\vert_{\mbox{\tiny f}} :=2 \sum_i s_i   \mathfrak{a}_i^\dagger  \mathfrak{a}_i     .
\label{bingo2}
\end{eqnarray}

\section{Quantum field theory: the  basic setup}

Let $M$ be a compact manifold and let $\Gamma$ be a general configuration space where each point is given by a field $\Theta$ on $M$. At this point we shall not specify exactly what type of field $\Theta$ is but leave that to the following sections. We are going to devise a general method of constructing a Hilbert space $L^2(\Gamma)$ via the Hilbert space construction presented in the previous section. To do this we first introduce a scalar product
$
\langle \cdot \vert \cdot \rangle_{\mbox{\tiny s}}
$
between fields in $\Gamma$ as well as a system $\{\xi_i  \}_{i\in\mathbb{N}}$, $\xi_i\in\Gamma$, with the properties
\begin{enumerate}
\item
that $\{\xi_i  \}_{i\in\mathbb{N}}$ is a real orthonormal basis with respect to $\langle \cdot \vert \cdot \rangle_{\mbox{\tiny s}}$, and
\item
that\footnote{In section 4.4 we find that for the commutator between the Bott-Dirac operator and a bounded field operator to be bounded we need the stronger condition $
\sum_i   \left\Vert {\xi_i} \right\Vert^2_\infty < \infty
$. See section 4.4 for details.}
$$
\sum_i s_i^{-1}  \left\Vert {\xi_i} \right\Vert^2_\infty < \infty
$$
with $\Vert \Theta \Vert^2_\infty = \sup_{m\in M} ( \Theta(m) , \Theta(m))$ where $(\cdot , \cdot)$ is an appropriate fiberwise scalar product that depends on the specifics of the configuration space $\Gamma$.
\end{enumerate}

The construction of $L^2(\Gamma)$ relies on the embedding
\begin{equation}
\Pi: \Gamma \rightarrow \mathbb{R}^\infty
\label{travlt}
\end{equation}
given by
$$
\Pi(\Theta) = (x_1,x_2, \ldots)
$$
where $x_i = \langle \Theta \vert \xi_i\rangle_{\mbox{\tiny s}}$. This embedding gives us a scalar product on functions on $\Gamma$
$$
\langle \eta(\Theta) \vert \zeta(\Theta) \rangle = \Big\langle  \eta\Big( \sum_i x_i \xi_i \Big) \Big\vert \zeta\Big(\sum_i x_i \xi_i \Big) \Big\rangle_{L^2(\mathbb{R}^\infty)}, 
$$
where $\eta$ and $\zeta$ are functions on $\Gamma$, that in turn allows us to define $L^2(\Gamma)$. With this construction we have a sequence of intermediate Hilbert spaces $L^2(\G_n)$ via the embeddings
$$
\G_n \ni \Theta=\sum_{i=1}^n x_i \xi_i \rightarrow (x_1,x_2,\ldots x_n)
$$
with the scalar products
\begin{eqnarray}
\langle \eta \vert \zeta \rangle_{\G_n} &=& \int_{\mathbb{R}^n} \overline{\eta(x_1\xi_1 + \ldots + x_n \xi_n)} \zeta (x_1\xi_1 + \ldots + x_n \xi_n) dx_1\ldots dx_n.
\end{eqnarray}
The Hilbert space $L^2(\G)$ is then the direct limit of these intermediate spaces
$$
L^2(\G)=\lim_{\rightarrow} L^2(\G_n).
$$
Furthermore, we shall later use the notation
$
\Lambda (\mathbb{R}^\infty ) = \Lambda^*\Gamma 
$
for the infinite-dimensional Clifford algebra.

Next let $U_{\oo}$ be the canonical translation operator given by
$$
U_\oo \eta(\Theta) = \eta(\Theta - \oo)
$$
where\footnote{Note that in the case of a gauge theory these translations are not given by gauge fields but by one-forms.} $\oo\in\Gamma$ and denote by $\mathrm{\bf Alg}(\Gamma)$ an algebra of bounded functions on the configuration space $\Gamma$. The exact form of $\mathrm{\bf Alg}(\Gamma)$ depends on the specifics of the configuration space $\Gamma$ but a general requirement is that a representation 
$$\rho:\mathrm{\bf Alg}(\Gamma)\rightarrow \cb(L^2(\Gamma)\otimes L^2(M))$$ 
exist. 
When this is the case we shall interpret the representation of the algebra generated by $\mathrm{\bf Alg}(\Gamma)$ and the translation operators $U_\oo$ as a kinematical sector of a quantum field theory over the configuration space $\Gamma$ and the square of the Bott-Dirac operator (\ref{udsigt}) as the Hamiltonian of the free theory.\\

In the following two sections we shall demonstrate this construction for two specific types of field theories, namely a real scalar field theory and a Yang-Mills theory, and we shall see that it enables us to define not only the free theories but also interacting ones.

\section{Scalar field theory}

In this section we will define $\mathrm{\bf Alg}(\Gamma)$ in the case of a real scalar field theory, and show that it has a representation on 
$L^2 (\Gamma )\otimes L^2(M)$. 

For the scalar field theory we will denote the configuration space by $\cs$. Furthermore a generic field will be denoted by $\phi$. 
The kind of elements we would like have in $\mathrm{\bf Alg}(\cs)$ are elements of the form $e^{i \phi}$. We will show that the representation is suitably strongly continuous, and thereby ensuring that we have self-adjoint operators like $\frac{d}{dt}e^{i \phi}|_{t=0}=i\phi$, and hence also operators like $\phi^4$. 

Like in the previous section $M$ is a compact manifold. Let $C^\infty_b(M\times \mathbb{R})$ be the bounded smooth  functions on $M\times \mathbb{R}$, where all the derivatives are also bounded (We actually only need the first three derivatives). Given $f\in C^\infty_b(M\times \mathbb{R})$ and $\phi \in \mathcal{S}$ we define an operator on $L^2(M)$ via 
$$(M_f(\phi ) \xi)(m)=f(m,\phi (m))\xi (m) .$$
We consider $M_f$ as an family of operators over $\mathcal{S}$.

 Following the general framework presented in the previous section we are now going to represent $C^\infty_b(M\times \mathbb{R})$ as operators on a Hilbert space $L^2(\cs)\otimes L^2(M)$, which will enable us to include the second ingredient, namely the translation operators on $\cs$. To construct $L^2(\cs)$ we first need to put a measure on $\mathcal{S}$. We follow the proceeding in the previous section. We choose a metric on $M$ and define the Sobolev scalar product
\begin{equation}
\langle \phi_1 \vert \phi_2 \rangle_{\mbox{\tiny s}} = \int_M \overline{(1+\tau_1\Delta^\sigma)\phi_1 }  (1+\tau_1\Delta^\sigma)\phi_2
\;,\quad \phi_1,\phi_2\in\cs
\label{iris}
\end{equation}
where $\sigma$ and $\tau_1$ are real numbers and where $\Delta$ is the scalar Laplace operator. Let $\{\xi_i\}_{i\in\mathbb{N}}$ be an orthonormal basis with respect to (\ref{iris}) where each $\xi_i$ is an eigenfunction of $\Delta$. 
Note that $\xi_i ({m}) = \frac{e_i({m}) }{1+ \tau_1 \lambda_i^\sigma}$ where $e_i({m}) $ is an eigenfunction of the Laplace operator with eigenvalue $\lambda_i$, i.e. $\Delta\a_i=\lambda_i\a_i$. 
Given a monotonously increasing sequence $\{s_i\}_{i\in\mathds{N}}$ we can choose $\sigma$ big enough such that
\begin{equation} \label{betingelse}
\sum_i   (s_i)^{-1}  \| \xi_i\|_\infty^2 <\infty .
\end{equation}
We then identify $\mathcal{S}$ with a subspace of $\mathbb{R}^{\infty}$ via the map
$$\mathcal{S} \ni  \phi =\sum_ix_i  {\xi_i} \to (x_1,x_2,\ldots )\in \mathbb{R}^\infty $$ 
and subsequently construct the Hilbert space $L^2 (\mathcal{S})$ and the action of $U_\omega $ thereon as outlined in the previous section. 
The translation operator $U_\oo$, $\oo\in \mathcal{S}$, act by
$$
U_\oo \eta(\phi) = \eta(\phi - \oo)\;,\quad \eta\in L^2(\cs).
$$
Like in \cite{Aastrup:2017vrm} the action of $U_\oo$ is strongly continuous and we therefore have infinitesimal operators 
$$
E_\omega=\frac{d}{dt} U_{t\oo}\vert_{t=0}.
$$
Finally, we build the full Hilbert space
$$
\ch_{\mbox{\tiny scalar}} = L^2(\cs)\otimes L^2(M),
$$
which carries an action of both the $M_f$ and $U_\oo$ operators.

\subsection{Proof of existence}

We need to show that the operators $M_f$ exist  on $\ch_{\mbox{\tiny scalar}}$. This question is similar to the case considered in \cite{Aastrup:2017vrm}.

We begin by showing 
\begin{thm} \label{bae}
 For each $f\in C^\infty_b (M\times \mathbb{R})$ and each  $m\in M$ the limit 
$$\lim_{n\to \infty}\int_{-\infty}^\infty \cdots \int_{-\infty}^\infty f\left( m , x_1\xi_1 (m)+\ldots   x_n\xi_n (m) \right)  e^{-\frac{s_1x_1^2+\ldots + s_nx_n^2}{\tau_2}}dx_1\cdots dx_n,$$
exists, i.e.
the expectation value in a given point $m$ of $M_f$ exists on the ground state in $L^2(\cs)$.

\end{thm}

\textit{Proof.} In a given point $(m,r)\in M\times \mathbb{R}$ we Taylor-expand $f$ in the $r$-direction:
$$f(m,x+r)=f(m,r)+xf'(m,r)+\frac{x^2}{2}f'' (m,r)+R_2(m,x+r),$$
with
$$ R_2(m,x+r) =\int_0^x \frac{t^2}{2}f''' (m,t+r)dt.$$
Here $f'=\frac{\partial f}{\partial r}$, $f''=\frac{\partial^2 f}{\partial^2 r}$, etc.
By asumption $f'''$ is bounded, let us say by  $B\tau^{-2}_2$. We hence  get the estimate 
$$ |R_2(m,x+r)| \leq |x|^3 B\tau^{-2}_2 .$$ 
We thus have 
\begin{eqnarray*}
\left(\frac{s}{\tau_2\pi}\right)^{1/2}\int_{-\infty}^\infty f(m,r+ax) e^{-\frac{s x^2}{\tau_2}} dx \hspace{-5cm}&&  \\ 
&=& \left(\frac{s}{\tau_2\pi}\right)^{1/2} \int_{-\infty}^\infty \Bigg( f(m,r)+axf'(m,r)\\
 &&
+\frac{(ax)^2}{2}f'' (m,r) +R_2(m,ax+r)  \Bigg) e^{-\frac{sx^2}{\tau_2}} dx\\
&=& f(m,r)+\left(\frac{s}{\tau_2\pi}\right)^{1/2} f'' (m,r)\int_{-\infty}^\infty \frac{(ax)^2}{2}e^{-\frac{s x^2}{\tau_2}} dx 
\\&&
+ \left(\frac{s}{\tau_2\pi}\right)^{1/2} \int_{-\infty}^\infty R_2(m,ax+r) e^{-\frac{sx^2}{\tau_2}}dx ,
\end{eqnarray*}   
where the first integral can be estimated by $\frac{\tau_2}{4s} |f'' (m,r)|a^2$ and the second integral by $s^{-2}Ba^3$. Putting in $a=\xi_k(m)$ and $s=s_k$ we get 
\begin{eqnarray*}
\left| f(m,r)-\left(\frac{s_k}{\tau_2\pi}\right)^{1/2}\int_{-\infty}^\infty f(m,r+x_k\xi_k) e^{-\frac{s_k x_k^2}{\tau_2}} dx \right| \hspace{-4cm}&&
\\&&
 \leq \frac{\tau_2}{4} |f''(m,r)| s_k^{-1} \|  \xi_k\|_\infty^2+B s_k^{-2} \|\xi_k\|_\infty^3 .
\end{eqnarray*}
Furthermore  $f''$ is bounded, let us say by $\frac{4C}{\tau_2}$, and hence
$$\left| f(m,r)-\left(\frac{s_k}{\tau_2\pi}\right)^{1/2}\int_{-\infty}^\infty f(m,r+x_k\xi_k) e^{-\frac{s_k x_k^2}{\tau_2}} dx \right|   \leq C s_k^{-1} \|\xi_k\|_\infty^2+ Bs_k^{-2}\|\xi_k\|_\infty^3 . $$ 
We thus have 
\begin{eqnarray}
\Bigg| \cn(n)\int_{-\infty}^\infty \cdots \int_{-\infty}^\infty f\left(m,\sum_{i=1}^n x_i\xi_i(m)\right)e^{-\frac{1}{\tau_2}\left(\sum_{i=1}^ns_ix_i^2\right)} dx_1\cdots d x_n \hspace{-10cm}&& \nn\\
&& - \cn(n+1)\int_{-\infty}^\infty \cdots \int_{-\infty}^\infty f\left(m,\sum_{i=1}^{n+1} x_i\xi_i(m)\right)e^{-\frac{1}{\tau_2}\left(\sum_{i=1}^{n+1}s_ix_i^2\right)} dx_1\cdots dx_{n+1} \Bigg| \nn\\
&& \leq Cs_{n+1}^{-1}\|e\xi_{n+1}\|_\infty^2+ Bs_{n+1}^{-2}\|\xi_{n+1}\|_\infty^3 .
\label{blob}
\end{eqnarray}
where $\cn(n)= \left(\frac{\p}{\tau_2}  \right)^{n/2}\P_{i=1}^n s_i^{1/2} $ is the normalisation factor for the $n$ Gaussian integrals.
The convergence of the expectation value of the $M_f$ operators on the ground state follows from (\ref{blob}) and (\ref{betingelse}). {\tiny\qed}
\\

It follows from the definition of the integrals, that the numerical value of the limit of the integrals is bounded by $\|f\|_\infty$. Like in \cite{Aastrup:2017vrm} it follows

\begin{thm}
For all $\xi ,\eta \in L^2(\cs )$ the limit
$$\langle \xi | M_{f(m,\cdot )} | \eta \rangle=\lim_{n\to \infty} \int_{\mathbb{R}^n} \overline{\xi (x_1,\ldots ,x_n) }f\left( m,\sum_{k=1}^nx_k\xi_k \right) \eta (x_1,\ldots ,x_n ) dx_1\cdots dx_n $$
exists.

\end{thm}

We now turn to strong continuity. We want to show that $\langle \xi | M_{f_k}(m) | \eta \rangle$ converges to $\langle \xi | M_{f}(m) | \eta \rangle$ when $f_k \to f$ in a suitable topology. The notion of convergence we choose is the following one: 
\begin{itemize}
\item The sequences $(f_k)$, $(f_k')$, $(f_k'')$ and $(f'''_k)$ are uniformly globally bounded, i.e. there exists a constant $K$ with $\|f_k\|_\infty ,\|f_k'\|_\infty,\|f_k''\|_\infty, \|f_k''' \|_\infty\leq K$.
\item $f_k\to f$, $f_k'\to f'$, $f_k''\to f''$ and $f'''_k \to f'''$ locally uniformly, i.e. uniformly on each compact subset of $M\times \mathbb{R}$.  
\end{itemize}

We have here chosen local uniformly, and not just uniformly. Uniform convergence is a too strict condition, since we will later use it for $f(m,x)=e^{itx}$, $t\to 0$.


Like before we start with convergence on the ground state.

\begin{thm}
Let $(f_k)$ be a sequence converging to $f$ in this topology. We have 
\begin{eqnarray*}  
&&\hspace{-1cm}\lim_{k\to \infty }\lim_{n\to \infty }\int_{\mathbb{R}^n}   \left( f\left( m ,\sum_{l=1}^n x_l\xi_1 (m) \right) - f_k\left( m , \sum_{l=1}^n x_l\xi_1 (m) \right) \right) e^{-\frac{1}{\tau_2}\left(\sum_{i=1}^ns_ix_i^2\right)}
\nn\\
&&\hspace{9cm}dx_1\cdots dx_n =0 
\end{eqnarray*}
\end{thm}

\textit{Proof.} The estimates in the proof of theorem \ref{bae}  show convergence if $f_k$, $f_k''$ and $f_k'''$ converges uniformly. We can however refine the argument slightly:

We first choose $n_0$ big enough with $$\sum_{n=n_0}^\infty s^{-1}_n\|\xi_n\|_\infty^2 <\varepsilon \quad \mbox{and} \quad \sum_{n=n_0}^\infty s^{-2}_n\|\xi_n\|_\infty^3 <\varepsilon.$$  
We put $$\tau_2^{-2}B_k=\sup |(f-f_k)'' (m,x)|\quad \mbox{and} \quad\frac{4C_k}{\tau_2}=\sup |(f-f_k)''' (m,x)|.$$  According to the estimates in the proof of theorem \ref{bae} we have 
\begin{eqnarray*} 
 \lim_{n\to \infty }\int_{\mathbb{R}^n}   \left( f\left( m ,\sum_{l=1}^n x_l\xi_1 (m) \right) - f_k\left( m , \sum_{l=1}^n x_l\xi_1 (m) \right) \right) e^{-\frac{1}{\tau_2}\left(\sum_{i=1}^ns_ix_i^2\right)}dx_1\cdots dx_n  
\end{eqnarray*}
equal to 
\begin{eqnarray*}  \int_{\mathbb{R}^{n_0}}   \left( f\left( m ,\sum_{l=1}^{n_0} x_l\xi_l (m) \right) - f_k\left( m , \sum_{l=1}^{n_0} x_l\xi_l (m) \right) \right) e^{-\frac{1}{\tau_2}\left(\sum_{i=1}^{n_0}s_ix_i^2\right)}dx_1\cdots dx_n  
\end{eqnarray*}
 up to an error of $2(B_k+C_k)\varepsilon$. Since $B_k $ and $C_k$ are bounded sequences we have that the error is $\varepsilon$ times some global factor. We thus only need to prove that   
\begin{eqnarray*}
&&\hspace{-1cm}   \lim_{k\to \infty} \int_{\mathbb{R}^{n_0}}   \left( f\left( m ,\sum_{l=1}^{n_0} x_l\xi_l (m) \right) - f_k\left( m , \sum_{l=1}^{n_0} x_l\xi_l (m) \right) \right) e^{-\frac{1}{\tau_2}\left(\sum_{i=1}^{n_0}s_ix_i^2\right)}\nn\\
&&\hspace{9cm}dx_1\cdots dx_n =0 
\end{eqnarray*}
This follows since $f_k $ converges locally uniformly to $f$, and we can, to a given error, choose a compact set, such that the integrals outside of this set is smaller that this error. This follows from the uniform  global boundedness of the sequence and the properties of the Gaussian integrals. {\tiny\qed}
\\

Like previously it is easy to extend this proof to  

\begin{thm}
Let $f_k\to f$. For all $\xi,\eta \in L^2(\cs)$ we have
 $$\lim_{k\to \infty}\langle \xi |  M_{f_k(m,\cdot )} | \eta \rangle =\langle \xi |  M_{f(m,\cdot )} | \eta \rangle$$
 We  thus have strong continuity in each point. 
\end{thm}

It follows from the argument for strong continuity that $m \to \langle \xi | M_{f}(m) | \eta \rangle$ is continuos. We thus have 

\begin{thm}
 For all $f\in C^\infty_b (M\times \mathbb{R})$ we  have  well defined bounded operators $M_f$ on $L^2(\cs) \times L^2( M )$ defined by 
$$ ((M_f)(\xi)) (m,\phi) =f(m, \phi (m) ) \xi (m,\phi ).$$
The action is strongly continuous. 

Especially we have an unbounded self-adjoint operator
$$i\phi : = \frac{d}{dt} M_{f_t} ,$$
where $f_t (m,r)= e^{itr}$. Consequently we also have any power of $\phi$ acting as self-adjoint unbounded operators on $L^2(\cs) \otimes L^2( M )$. 

\end{thm}

\subsection{Comparison with a canonically quantised real scalar field}

We are now going to compare this construction to the case of a canonically quantised real scalar field.
In the following we let $M=\mathds{T}^3$, the 3-torus.

Consider first the infinitesimal translation operators $E_\oo = \frac{d}{ds} U_{s \oo}\vert_{s=0}$. Expanding $\oo$ in the Sobolev eigenvectors these can be written as
$$
E_\oo = \sum_i \oo_i E_{\xi_i},\quad \oo=\sum_i \oo_i \xi_i.
$$
Note also that there exit a canonical operator
$$
E(m) = \sum_i \xi_i(m) E_{\xi_i} = \sum_i \frac{\pa}{\pa x_i} \xi_i (m)
$$
where we used the identification $E_{\xi_i}= \frac{\pa}{\pa x_i}$. With this type of operators
we can form an alternative representation given by the linear combinations
\begin{eqnarray}
{\phi}' (m)  &=&\frac{1}{\sqrt{2}}  \sum_{i=1}^\infty\Big[ x_i \left( \xi_i(m) + \xi_i(-m)  \right) +  \frac{\tau_2}{s_i}\frac{\pa}{\pa x_i} \left(  \xi_i(m) - \xi_i(-m) \right)      \Big]
\nn\\
&=&   \sum_{i=1}^\infty \frac{1}{\sqrt{2s_i}} \left(  q_i \xi_i(m) + q_i^\dagger \xi_i(-m)  \right),
\nn\\
\p (m)  &=& \frac{1}{\sqrt{2}}  \sum_{i=1}^\infty \Big[   s_i x_i \left( \xi_i(m) + \xi_i(-m)  \right) -  {\tau_2}\frac{\pa}{\pa x_i} \left(  \xi_i(m) - \xi_i(-m) \right)      \Big]
\nn\\
&=&  \sum_{i=1}^\infty \sqrt{\frac{s_i}{2}} \left(  q_i \xi_i(m) - q_i^\dagger \xi_i(-m)  \right),
\label{B1}
\end{eqnarray}
which shall shortly be seen to provide a connection to the canonical quantization of a real scalar field.

But before we get that far let us first specify the parameters $\{s_i\}_{i\in\mathbb{N}}$ by $s_i = \sqrt{p_i^2 + {\bf m}^2}\equiv \oo_{p_i}$, where ${\bf m}$ is a real constant that plays the role of a mass and where $\{p_i\}_{i\in\mathbb{N}}$ is a sequence of parameters that plays the role of a momentum.
With this operator $B^2\big\vert_{\mbox{\tiny b}}$ in  (\ref{bingo2}) has the form
\begin{equation}
B^2\big\vert_{\mbox{\tiny b}} = \sum_i  {\oo_{p_i}} q_i^\dagger q_i .
\label{B2}
\end{equation}

Let us now compare this construction and in particular equations (\ref{B1}) and (\ref{B2}) with a canonical quantisation of a free, real scalar field. We therefore let $M=\mathbb{R}^3$ and denote by $\P(m)$ the conjugate to the real scalar field $\Phi(m)$. As is custom we expand $\Phi({m})$ and $\P({m})$ in plane waves according to
\begin{eqnarray}
\Phi({m})  &= &\int \frac{d^3p}{(2\p)^3} \frac{1}{\sqrt{2\omega_{{p}}}}\left( a_{{p}} e^{i {p}\cdot {m}}  + a^{\dagger}_{{p}} e^{-i {p}\cdot {m}}            \right)
\nn\\
\P({m}) &=& \int \frac{d^3p}{(2\p)^3}(-i) \sqrt{  \frac{\omega_{{p}}}{2} }\left( a_{{p}} e^{i {p}\cdot {m}}  - a^{\dagger}_{{p}} e^{-i {p}\cdot {m}}            \right)
\label{C1}
\end{eqnarray}
where $\omega_{{p}}=\sqrt{{p}^2+{\bf m}^2}$ and where $a_{{p}}$ and $a^\dagger_{{p}}$ are the creation and annihilation operators (indexed by the 3-momentum ${p}$) that act in a corresponding Fock space. Also, the Hamilton operator for the free scalar field is given by
\begin{equation}
H_{\mbox{\tiny free}} = \int \frac{d^3p}{(2\p)^3} \oo_{{p}} a_{{p}} ^\dagger a_{{p}} 
\label{C2}
\end{equation}

Already here we see a clear resemblance between the embedding (\ref{B1}) and the plane wave expansion (\ref{C1}) and between the square of the Bott-Dirac operator (\ref{B2}) and the Hamiltonian for the free scalar field (\ref{C2}).
If we take a limit where $M$ goes from being $\mathds{T}^3$ to approaching $\mathbb{R}^3$:
\begin{eqnarray}
p_i &\longrightarrow &{p}
\nn\\
\sum_i& \longrightarrow &\int \frac{d^3p}{(2\p)^3} 
\nn
\end{eqnarray}
in which case the Sobolev eigenvectors $\xi_i$ can be written
$$
\xi_i(m)  \longrightarrow \frac{e^{- i p\cdot m}}{1+ \tau_1\lambda_i^\sigma} ,
$$
then it is clear that the framework we have presented in this section is, in the local limit $\tau_1\rightarrow 0$, identical to that of a canonically quantised free scalar field. 
Note that the Hilbert space representation, which we have constructed in this section, ceases to exist precisely in the limit $\tau_1\rightarrow 0$, which is to be expected as no such representation exist for ordinary (interacting) perturbative quantum field theory. We shall discuss this further in section \ref{locality}.\\

To construct the Hamilton of an interacting theory we need to consider a Hamilton operator of the form 
$$H|_{\mbox{\tiny b}}+H_{\mbox{\tiny int}} ,$$
where for example $H_{\mbox{\tiny int}}=\phi^4$. We know that both $H|_{\mbox{\tiny b}}$ and $H_{\mbox{\tiny int}}$ exists as self-adjoint unbounded operators on $\ch_{\mbox{\tiny scalar}}$. Strictly speaking we have not proved that their sum exists but we are certain this will not be hard to prove. We think, however, that this should come with a more detailed analysis of the domains of the operators, as well as the development of a pseudo-differential calculus. For instance the natural Sobolev spaces should be given by
$$H^k_{\mbox{\tiny Sobolev}}(\cs )= \hbox{Domain of } H|_{\mbox{\tiny b}}^{\frac{k}{2}}.$$  

Another point is that we have in the case of the scalar theory chosen a rather minimal algebra. The chosen algebra allows for operators like $\phi^n$, but does not allow derivatives in $\phi$ for instance. The situation with the Holonomy-Diffeomorphism algebra, which we shall discuss in the next section, is different, since this algebra contain enough information to separate each gauge-orbit.   \\

It is illustrative to rewrite the expectation value of the operator $M_f$ in $L^2(\cs)$ with the short-hand notation
\begin{equation}
\left\langle \eta_{\mathrm{\bf gs}} (\phi) \left\vert M_f(\phi ) \right\vert \eta_{\mathrm{\bf gs}}(\phi)  \right\rangle_{L^2(\cs)} (m)
=
\int_\cs d\phi f(\phi(m)) \exp \left( - \Vert  {\oo_p}\phi \Vert^2_{\mbox{\tiny s}}  \right),
\label{shorthand}
\end{equation}
which has the form of a functional integral where the Sobolev norm $\Vert\cdot\Vert_{\mbox{\tiny s}} $ plays the role of a weight. With this heuristic notation it becomes clear that the functional integral is dominated by those field configurations, which have a small Sobolev norm and that field configurations, that have a large Sobolev norm, are dampened. In particular, this means that singular field configurations, i.e. those that are localised in a single point, will have {\it zero} weight in this integral. To see this we simply note that the Sobolev norm dominates the supremum norm\footnote{This statement depends on the choice of $\sigma$ in (\ref{iris}).} \cite{Nirenberg} and since the supremum norm of a delta function is infinite so is the Sobolev norm and hence the exponential factor in (\ref{shorthand}) will be equal to zero. This illustrates that the quantum field theories, which we are presenting in this paper, are only local up to the scale $\tau_1$.

Another way to see this non-local aspect is by noting that the plane waves in the operator expansion (\ref{C1}) used in ordinary quantum field theory and which corresponds to a point-localisation due to $\int d^3p e^{ip\cdot m}\sim\d^{(3)}(m)$, are in (\ref{B1}) replaced by the Sobolev eigenvectors $\xi_i$, which only correspond to a point-wise localisation up to a correction at the order of the scale $\tau_1$.\\

Before we end this section let us briefly consider again the embedding (\ref{B1}). The reason why we write down this particular combination is that it matches the corresponding plane wave expansion in perturbative quantum field theory. The question is, however, if there exist a deeper reason for this structure. 

Note first that when $\{\x_i\}_{i\in\mathds{N}}$ are given by plane waves (when we set $M=\mathds{T}^3)$, then $ \xi_i (m)\pm \xi_i(-m)$ are their real and imaginary parts respectively. Thus, when '$x_i$' appears in (\ref{B1}) only in combination with $ \xi_i (m) + \xi_i(-m)$, then it ensures that the expansion $\sum_i x_i \xi_i$ is well defined, i.e. that the vectors in the expansion are real. We have assumed that $\xi_i$ are real eigenfunctions for this reason but it appears that quantum field theory has already taken this into account. 

Let us for now no longer assume that $\{\xi_i\}_{i\in\mathds{N}}$ are real and rewrite (\ref{B1}) in the form
\begin{equation}
\left(
\begin{array}{c}
\phi'\\
\p
\end{array}
\right) =\sum_{i=1}^\infty
\frac{1}{\sqrt{2}} \left(
\begin{array}{cc}
1& s_i^{-1}\\
s_i &- 1
\end{array}
\right) 
\left(
\begin{array}{c}
x_i \mbox{Re}( \xi_i)\\
\tau_2 i \frac{\pa}{\pa x_i} \mbox{Im}(\xi_i)
\end{array}
\right) 
\end{equation}
where
\begin{eqnarray}
\mbox{Re}(\xi_i) = \xi_i + \xi_i^* ,\quad \mbox{Im}(\xi_i) = \xi_i - \xi_i^*.
\nn
\end{eqnarray}
We then find the operators
$$
J_i = \frac{1}{\sqrt{2}}\left(
\begin{array}{cc}
1& s_i^{-1}\\
s_i &- 1
\end{array}
\right) ,\quad J_i^2 = \mathds{1}.
$$
It is an interesting question whether there exist a mathematical explanation for this particular form.

\subsection{The signature of the metric}

Note that with the construction, that we have presented so far, the space-time metric is not an input, as is the case in most other approaches to quantum field theory. Indeed, the only geometrical input is the metric $g$ on the 3-dimensional manifold $M$. This raises the interesting question what 4-metric will emerge from the construction with a time-evolution given by the Hamilton operator and whether the quantum theory will be covariant with respect to this metric. 

One piece of information about the emergent 4-dimensional metric can be determined immediately, namely its signature. The correspondence between our construction and a canonically quantised real scalar field, which we have demonstrated in this section, is compatible with a 4-dimensional metric that has a Minkowski signature. Since the Bott-Dirac operator is a canonical structure it does not seem possible to incorporate any other signature.

\subsection{A semi-finite spectral triple over $\cs$}

So far we have identified three basic ingredients from which we have built a scalar quantum field theory. These are the Bott-Dirac operator $B$, the algebra $C^\infty_b(M\times \mathbb{R})$ and the Hilbert space $L^2(\cs)\otimes \Lambda^*\cs\otimes L^2(M)$.
Since we know that $B$ has compact resolvent modulo $\Lambda^*\cs$ when acting on $L^2(\cs) \otimes  \Lambda^*\cs $ the question arises whether the commutator $[B,M_a]$, $a\in C^\infty_b(M\times \mathbb{R})$, is bounded. The commutator can be computed as
\begin{eqnarray*}
&&[B,M_a](m,x_1\xi_1 (m)+x_2\xi_2 (m)+\ldots )\\
&&\hspace{1cm}
 =\sum_{k=1}^\infty \bar{c}_k\xi_k(m) a'(m,x_1\xi_1 (m)+x_2\xi_2 (m)+\ldots ) .
\end{eqnarray*}
We therefore see that the commutator exist and is bounded when 
\begin{equation} \label{con-s}
\sum_{k=1}^\infty \| \xi_k \|^2_{\infty} < \infty .
\end{equation}
The operator $B$ does not have compact resolvent modulo $\Lambda^*\cs $ when it acts on $L^2(\cs) \otimes  \Lambda^*\cs \otimes L^2 (M) $, due to the $L^2 (M) $ factor. We can however repair this: The algebra $C_b^\infty (M,\mathbb{R})$ is acting on $L^2(\cs \times M)=L^2 (\cs)\otimes L^2 ( M)$ as an algebra of functions. We can therefore consider the trace 
$$Tr_{\cs} \otimes \tau \otimes \tau_g  , $$
where  $Tr_{\cs}$ is the ordinary trace on $\mathbb{B} (L^2 (\cs))$, $\tau$ is the finite normalised trace on $\Lambda^*\cs$, and  $\tau_g$ is the finite trace on functions  on $M$ given by 
$$\tau_g ( f )=\int_M f(m) dg (m) .$$ 
With this trace the triple 
\begin{equation}
(B,C^\infty_b(M\times \mathbb{R}),L^2(\cs)\otimes \Lambda^*\cs \otimes L^2 (M)   )
\label{tripleA}
\end{equation}
becomes a semi-finite spectral triple when the condition (\ref{con-s}) is fulfilled. 
This implies that a scalar quantum field theory can be understood as a geometrical construction over the configuration space $\cs$ of scalar field configurations.

\section{Yang-Mills theory}

We are now going to construct a quantum Yang-Mills theory. Let therefore $M$ be a compact manifold and let $\ca$ be a configuration space of gauge connections that takes values in the Lie-algebra of a compact gauge group $G$ and let $\mathrm{\bf Alg}(\ca)= \mathbf{HD}(M)$, where $\mathbf{HD}(M)$ is an algebra generated by holonomy-diffeomorphisms as will be described next and which was first defined in \cite{Aastrup:2012vq,AGnew}. 


A holonomy-diffeomorphism $f e^X\in \mathbf{HD}(M)$, where $f\in C^\infty(M)$, is a parallel transport along the flow $t\to \exp_t(X)$ of a vector field $X$.  
To see how this works we first let $\gamma$ be the path
$$\gamma (t)=\exp_{t} (X) (m) $$
running from $m$ to $m'=\exp_1 (X)(m)$. Given a connection $\nabla$ that takes values in a $n$-dimensional representation of the Lie-algebra $\mathfrak{g}$ of $G$  we then define a map
$$e^X_\nabla :L^2 (M )\otimes \mathbb{C}^n \to L^2 (M )\otimes \mathbb{C}^n$$
via the holonomy along the flow of $X$
\begin{equation}
  (e^X_\nabla \xi )(m')=    \hbox{Hol}(\gamma, \nabla) \xi (m)   ,
  \label{chopin1}
 \end{equation}
where $\xi\in L^2(M,\mathbb{C}^n)$ and where $\hbox{Hol}(\gamma, \nabla)$ denotes the holonomy of $\nabla$ along $\gamma$. This map gives rise to an operator valued function on the configuration space $\ca$ of $G$-connections via
\begin{equation}
\ca \ni \nabla \to e^X_\nabla  ,
\nn
\end{equation}
which we denote by $e^X$. For a function $f\in C^\infty (M)$ we get another operator valued function $fe^X$ on $\ca$, which we call a holonomy-diffeomorphisms\footnote{The holonomy-diffeomorphisms, as presented here, are not a priori unitary, but by multiplying with a factor that counters the possible change in volume in (\ref{chopin1}) one can make them unitary, see \cite{AGnew}.}.

Furthermore, a $\mathfrak{g}$ valued one-form $\oo$ induces a transformation on $\ca$ and therefore an operator $U_\omega $ on functions on $\ca$ via   
$$
U_\omega (\xi )(\nabla) = \xi (\nabla - \omega) ,
$$ 
which gives us the quantum holonomy-diffeomorphism algebra, denoted $\mathbf{QHD}(M)$, as the algebra generated by $\mathbf{HD}(M)$ and all the $U_\oo$ operators (see \cite{Aastrup:2014ppa}).

To obtain a representation of the $\mathbf{QHD}(M)$ algebra we let 
$\langle \cdot\vert\cdot\rangle_{\mbox{\tiny s}} $ denote the Sobolev norm on $\OO^1(M\otimes\mathfrak{g})$, which has the form
\begin{equation}
\langle \omega_1\vert\omega_2\rangle_{\mbox{\tiny s}} 
:=
\int_M dx \big( (1+ \tau_1\Delta^{\sigma})\omega_1  , (1+  \tau_1\Delta^{\sigma})\omega_2  \big)_{T_x^*M\otimes \mathbb{C}^n}
\label{sob}
\end{equation}
where the Hodge-Laplace operator $\Delta$ and the  inner product  $(,)_{T_x^*M\otimes \mathbb{C}^n}$ on $T_x^*M\otimes \mathbb{C}^n$ depend on a metric g and where $\tau_1$ and $\sigma$ are positive constants. Also, we choose an $n$-dimensional representation of $\mathfrak{g}$.

Next, denote by $\{\b_i\}_{i\in\mathbb{N}}$ an orthonormal basis of $\OO^1(M\otimes\mathfrak{g})$ with respect to the scalar product (\ref{sob}).
With this we can construct a space $L^2(\ca)$ as an inductive limit over intermediate spaces $L^2(\ca_n)$ with an inner product given by
\begin{eqnarray}
\langle \eta \vert \zeta \rangle_{\ca_n} &=& \int_{\mathbb{R}^n} \overline{\eta(x_1\b_1 + \ldots + x_n \b_n)} \zeta (x_1\b_1 + \ldots + x_n \b_n) dx_1\ldots dx_n
\end{eqnarray}
where $\eta$ and $\zeta$ are elements in $L^2(\ca)$, as explained in section 3. Finally, we define the Hilbert space 
\begin{equation}
\ch_{\mbox{\bf\tiny YM}}= L^2(\ca)\otimes L^2(M, \mathbb{C}^n)
\label{ymm}
\end{equation}
in which we then construct the following representation of the $\mathbf{QHD}(M)$.
First, given a smooth one-form $\oo\in\OO^1(M,\mathfrak{g})$ we write $\oo =\sum \oo_i \b_i$. The operator $U_\chi$  acts by translation in $L^2(\ca)$, i.e. 
\begin{eqnarray}
U_{\oo}(\eta) (\omega)&=&U_{\oo}(\eta) (x_1 \b_1+x_2 \b_2+ \ldots)
\nn\\
&=&  \eta ( (x_1+\oo_1)\b_1+(x_2+\oo_2)\b_2+ \ldots)  
\label{rep1}
\end{eqnarray}
with $\eta\in L^2(\ca)$. Next, we let $f e^X\in \mathbf{HD}(M)$ be a holonomy-diffeomorphism and $\Psi(\omega,m)=\eta(\omega)\otimes \psi(m)\in\ch_{\mbox{\tiny\bf YM}}$ where $\psi(m)\in L^2(M)\otimes \mathbb{C}^n$. We write
\begin{equation}
f e^X \Psi(\omega,m') =  f(m) \eta(\omega) Hol(\gamma, \omega) \psi(m)  
\label{rep2}
\end{equation}
where $\gamma$ is again the path generated by the vector field $X$ with $m'=\exp_1(X)(m)$.

\begin{thm}

Equations (\ref{rep1}) and (\ref{rep2}) gives a strongly continuous Hilbert space representation of the $\mathbf{QHD}(M)$ algebra in $\ch_{\mbox{\bf\tiny YM}}$.

\end{thm}

\textit{Proof.} 
  In \cite{Aastrup:2017vrm} we prove that (\ref{rep1}) and (\ref{rep2}) give rise to a strongly continuous Hilbert space representation of the $\mathbf{QHD}(M)$ algebra in the special case where $s_i = 1$ for all $ i\in\mathbb{N}$. This proof can be straight forwardly adopted to the case where $\{s_i\}_{i\in\mathds{N}}$ is a monotonously increasing sequence and we leave it to the reader to check this.
 {\tiny\qed}

\subsection{Comparison with a canonically quantised gauge field}

Just as we did for the scalar field  we now let $M=\mathds{T}^3$, the 3-torus, and then form a linear combination
\begin{eqnarray}
{ A} (m)  
&=&\sum_i \frac{1}{\sqrt{2s_i}} \left(  q_i \b_i(m) + q_i^\dagger \b_i(-m)  \right),
\nn\\
{ E} (m)  
&=& \sum_i \sqrt{\frac{s_i}{2}} \left(  q_i \b_i(m) - q_i^\dagger \b_i(-m)  \right),
\label{D1}
\end{eqnarray}
which is similar to (\ref{B1}) but where we must keep in mind that '$i$' is a multi-index, that includes also Lie-algebra and vector degrees of freedom. To clarify this let us split the summations in (\ref{D1}) up by separating out the Lie-algebra and spatial part:
\begin{eqnarray}
A^a (m)  
&=&\sum_{k,r} \frac{1}{\sqrt{2s_k}} \frac{\e_k^r }{1+\tau_1\lambda_k^\sigma}\left(  q^a_{k,r} e^{ik\cdot m} + q_{k,r} ^{a\dagger} e^{-ik\cdot m}  \right),
\nn\\
E^a (m)  
&=& \sum_{k,r} \sqrt{\frac{s_k}{2}} \frac{\e_k^r }{1+\tau_1\lambda_k^\sigma} \left(  q^a_{k,r} e^{ik\cdot m} - q_{k,r}^{a\dagger} e^{-ik\cdot m}  \right),
\label{D2}
\end{eqnarray}
where $r\in\{1,2,3\}$ are the spatial indices and '$a$' a Lie-algebra index. We have also assumed that the sequence $\{s_n\}_{n\in\mathds{N}}$ only depends on the index $k$.

Let us this time fix the parameters with $s_k =\oo_{p_k}=  \vert p_k \vert$, which gives $B^2\big\vert_{\mbox{\tiny b}}$ in  (\ref{bingo2}) the form
\begin{equation}
B^2\big\vert_{\mbox{\tiny b}} = \sum_{k,r}    {\vert p_k\vert} q_{k,r}^\dagger q_{k,r}
\label{D3}
\end{equation}

Compare this construction to perturbative quantum field theory of a general gauge field ${\bf A}(m)$ and its conjugate ${\bf E}(m)$ on $M=\mathbb{R}^3$ in the Coulomb gauge \cite{Campagnari:2009km}. Within the framework of canonical quantisation these fields are expanded according to
\begin{eqnarray}
{\bf A}({m}) &=& \int \frac{d^3p}{(2\p)^3} \frac{1}{\sqrt{2\vert{p}\vert}} \sum_{r=1}^2 {\e}_r ({p}) \left(  a^r_{{p}} e^{i {p}\cdot {m}}   +  a^{r\dagger}_{{p}} e^{-i {p}\cdot {m}}   \right)
\nn\\
{\bf E}({m}) &=& \int \frac{d^3p}{(2\p)^3} (-i)\sqrt{\frac{\vert{p}\vert}{ 2 }} \sum_{r=1}^2 {\e}_r ({p}) \left(  a^r_{{p}} e^{i {p}\cdot {m}}   -  a^{r\dagger}_{{p}} e^{-i {p}\cdot {m}}   \right)
\label{F1}
\end{eqnarray}
where $\e_r(p)$ is a set of polarisation vectors and where $a_p$ and $a_p^\dagger$ are creation and annihilation operators acting in a corresponding Fock space. Note that these are Lie-algebra valued. Also, the Hamiltonian of the free theory is in the Coulomb gauge given by \cite{Campagnari:2009km}
$$
H_{\mbox{\tiny\bf free}}  =  \int \frac{d^3p}{(2\p)^3} \vert {p}\vert \sum_{r=1}^2 \mbox{Tr}_{\mathfrak{g}} a^{r\dagger}_{{p}} a^r_{{p}} 
$$
where $ \mbox{Tr}_{\mathfrak{g}}$ denotes a trace over the Lie-algebra ${\mathfrak{g}}$.

We see that the construction of a Bott-Dirac operator interacting with a representation of the $\mathbf{HD}(M)$ algebra coincides with the free sector of a canonically quantised gauge field in the Coulomb gauge when we take the flat and local limits $M\rightarrow \mathbb{R}^3$, $\tau_1\rightarrow 0$. The only discrepancy is that the sum over $r$ in (\ref{F1}) runs only over transversal degrees of freedom where as the sum in (\ref{D2}) runs over all three spatial directions. This is due to the fact that we have not restricted the degrees of freedom in the embedding 
$$\ca\ni\oo=\sum_i \oo_i \b_i \rightarrow (x_1,x_2,...)$$ 
of the configuration space $\ca$ into $\mathbb{R}^\infty$ to only include transversal degrees of freedom. This can be straight forwardly done, however.

Thus we conclude that also in the case of a gauge theory does the general framework of non-perturbative quantum field theory, that we have presented, coincide with that of a canonically quantised gauge field in the flat and local limit and with the free Hamiltonian given by the square of the Bott-Dirac operator.

It is remarkable that it is the same Bott-Dirac operator, that gives rise to the Hamiltonians in {\it both} the case of a free scalar field and in the case of the free sector of a gauge field. In the next section we will see that the Bott-Dirac operator also gives rise to the Hamiltonian of a quantised fermionic field.

Note again that the signature of the emergent 4-dimensional metric, that is compatible with the above analysis, will have a Minkowski signature.\\

Next let us briefly consider also the full Yang-Mills Hamiltonian, which can be written
\begin{equation}
H_{\mbox{\tiny YM}} = \frac{1}{2}\int d^3x \left( (E^a_\m)^2 + (B^a_\m)^2  \right)
\label{bee}
\end{equation}
where $E$ is again the conjugate field to the gauge field $A$ and where $B_\m=  \e_\m^{\;\;\n\s} F_{\n\s}$ with $F$ being the field strength tensor of $A$.
In \cite{Aastrup:2017vrm} we showed that the operators $ (B^a_\m)^2$ and $(E^a_\m)^2$ can be constructed within our framework, where they will be local only up to a correction at the order of the scale $\tau_1$.

\subsection{A semi-finite spectral triple over $\ca$}

We will here consider the $*$-subalgebra $\mathbf{HD}(M)|_{\mbox{\tiny{loops}}}$ of $\mathbf{HD}(M)$ generated by the closed flows. The representation of this algebra generated by a given connection is an algebra of matrix valued functions over $M$. We can therefore, like in the case of the scalar theory, consider the trace 
$$Tr_{\cs} \otimes \tau \otimes \tau_g  , $$
but where 
$$\tau_g ( f )=\int_M Tr_{\mathbb{C}^n}\left( f(m) \right) dg (m) .$$

Just as it was the case for the scalar theory we note that we have a semi-finite spectral triple over $\ca$ consisting of
$$
(B, \mathbf{HD}(M)|_{\mbox{\tiny{loops}}}, \ch'_{\mbox{\bf\tiny YM}}).
$$
under the condition (\ref{con-s}) and where $\ch'_{\mbox{\bf\tiny YM}}=\ch_{\mbox{\bf\tiny YM}}\otimes \Lambda^*\ca$. 

We have chosen to consider the $*$-sub-algebra $\mathbf{HD}(M)|_{\mbox{\tiny{loops}}}$, since it is not clear how one is to define a trace on the $L^2(M, \mathbb{C}^n))$ part of the Hilbert space with the full $\mathbf{HD}(M)$ algebra.

\section{Fermionic quantum field theory}

Since our discussion in the previous sections has been concerned with bosonic quantum field theory only the question arises whether fermionic quantum field theory has a place in our framework as well. This is the topic of the following discussion, where we will show that this is indeed the case.\\

So far we have found that the free Hamiltonian of a bosonic field theory is given by the square of the Bott-Dirac operator (\ref{udsigt}) acting in the Hilbert space $L^2(\mathbb{R}^\infty)$. The Bott-Dirac operator itself acts, however, in the Hilbert space $L^2(\mathbb{R}^\infty)\otimes \Lambda^*\mathbb{R}^\infty$ and the square of the Bott-Dirac operator involves besides the harmonic oscillator also the operator $B^2\big\vert_{\mbox{\tiny f}}$, which acts in $\Lambda^*\mathbb{R}^\infty$ only. Thus, the additional factor $\Lambda^*\mathbb{R}^\infty$ in the Hilbert space involves structure which, as we shall see, amounts precise to a fermionic quantum field theory.

Let us begin with a spinor $\psi$ and its conjugate $i \psi^\dagger$ on $M=\mathbb{R}^3$ and its canonical quantisation in terms of plane waves
\begin{eqnarray}
\psi(m) &=& \int \frac{d^3p}{(2\p)^3}  \sum_{s=1}^2 \frac{1}{\sqrt{2 \oo_p}} \left[ b_p^s u^s(p) e^{-ip\cdot x}  +  c_p^{s\dagger} v^s(p) e^{ip\cdot x}    \right]
\nn\\
\psi^\dagger(m) &=& \int \frac{d^3p}{(2\p)^3}  \sum_{s=1}^2 \frac{1}{\sqrt{2 \oo_p}}  \left[ b_p^{s\dagger} u^{s\dagger}(p) e^{ip\cdot x}  +  c_p^{s} v^{s\dagger}(p) e^{-ip\cdot x}    \right]
\label{little}
\end{eqnarray}
where $u$ and $v$ are spinors and $(b,b^\dagger)$ and $(c,c^\dagger)$ are the associated pairs of creation and annihilation operators that satisfy the anti-commutation relations
\begin{eqnarray}
\{ b_p^r, b_q^{s\dagger} \} = \{ c_p^r, c_q^{s\dagger} \}  = (2\p)^3 \d^{rs}\d^{(3)}(p-q)
\nn
\end{eqnarray}
while all other anti-commutators vanish
\begin{eqnarray}
\{ b_p^r, b_q^{s} \} = \{ b_p^{r\dagger}, b_q^{s\dagger} \} = \{ c_p^r, c_q^{s} \} = \{ c_p^{r\dagger}, c_q^{s\dagger} \} =\ldots = 0.
\end{eqnarray}
Also, the Hamilton operator for a quantised spinor field reads
\begin{equation}
H_{\mbox{\tiny spinor}} = \int \frac{d^3p}{(2\p)^3}  { \oo_p} \left(  b_p^{s\dagger} b_p^{s}  +  c_p^{s\dagger} c_p^{s}   \right).
\label{queen}
\end{equation}

Now, we would like to repeat the line of interpretation employed in the previous sections, where the plane waves were viewed as flat-space and local limits of eigenfunctions of a Laplace operator and the momentum integrals were viewed as limits of sums over these eigenfunctions.
To this end note first that the operators $(\mathfrak{a}_i,\mathfrak{a}^\dagger_i)$ in (\ref{rolignu}) and the exterior algebra $\Lambda^*\mathbb{R}^\infty$ gives us precisely the CAR algebra of creation and annihilation operators. Furthermore, note that the square of the Bott-Dirac operator gives us the fermionic operator
$$
\frac{1}{2}B^2\big\vert_{\mbox{\tiny f}} := \sum_i \oo_i   \mathfrak{a}_i^\dagger  \mathfrak{a}_i   
$$
that has precisely the form of the fermionic Hamilton (\ref{queen}) when the limits mentioned above are taken into consideration and if we identify again $s_i=\oo_i$. Thus, we see that the construction, that we have presented in this paper, naturally includes fermionic quantum field theories too.

There is a caveat, however, which is that the number of degrees of freedom in the fermionic sector has to match that of the bosonic sector. Whereas it makes sense to consider a bosonic field on its own  -- we can just consider the Hilbert space $L^2(\mathbb{R}^\infty)$ without the infinite-dimensional exterior algebra  -- a fermionic field will in this framework always come together with a bosonic field and have the same number of degrees of freedom. For a scalar field theory this appears to be a problem due to the spin-statistics theorem, but a gauge field, with two transversal degrees of freedom, can match that of a 2-spinor as long as the gauge field and the spinor field transform in the same representation of the gauge group.

\subsection{The Bott-Dirac commutator}
\label{flu}

Let us once more consider a real scalar field and its embedding into $\mathbb{R}^\infty$ in (\ref{B1})  and let us consider the commutator between the Bott-Dirac operator and the field $\phi'(m)$ on $M=\mathds{T}^3$, i.e.  $\phi'(m)= \sum_i \frac{1}{\sqrt{2s_i}} (q_i \xi_i(m)  +  q_i^\dagger \xi_i(-m))$, which reads
\begin{eqnarray}
[B, \phi'(m)]  =  \sum_i \frac{1}{\sqrt{2}} (\mathfrak{a}^\dagger_i \xi_i(m)  +  \mathfrak{a}_i \xi_i(-m))
\label{bott-comm}
\end{eqnarray}
Now, if we for a moment ignore the spinors $u$ and $v$ in equation (\ref{little}) and once more view the operator expansions in canonical quantisation as a flat and local limit of an expansion in terms of Sobolev eigenvectors $\xi_i$, i.e.
\begin{eqnarray}
\sum_i &\longleftrightarrow& \int \frac{d^3 p}{(2\p)^3}
\nn\\
\xi_i (m) &\longleftrightarrow& e^{i m\cdot p},
\nn\\
s_i  &\longleftrightarrow& \oo_p
\label{corre}
\end{eqnarray}
then we see a clear resemblance between the commutator (\ref{bott-comm}) and the conjugate fermionic field operator $\psi^\dagger$ in  (\ref{little}), where the operators $(\mathfrak{a}^\dagger_i,\mathfrak{a}_i)$ generate the CAR algebra. Schematically, we have the relations
\begin{eqnarray}
[B, \mbox{"boson"}] &=& \mbox{"fermion"} , \nn\\ \{B,\mbox{"fermion"}\} &=&\mbox{"boson"}\nn
\end{eqnarray}
which corresponds to the relation $[ d, f] = df$, where the Bott-Dirac operator is the differential operator, the "boson" a zero-form and the "fermion" a one-form. The only serious discrepancy is that the "fermion" that the Bott-Dirac operator generates is, in this case, a scalar, which seems to be in violation with the spin-statistics theorem. Note also that the allocation of factors of $s_i$ in (\ref{bott-comm}) is different from that in (\ref{little}).

If we instead consider a gauge field and its expansion (\ref{D1}) 
\begin{eqnarray}
[B, A(m)]  =   \sum_i \frac{1}{\sqrt{2}} (\mathfrak{a}_i \b_i(m)  +  \mathfrak{a}_i^\dagger \b_i(-m))  = \tilde{\psi}(m)
\label{bott}
\end{eqnarray}
then we find $\tilde{\psi}(m)$, which is a fermionic field that has, once we take the correspondence  (\ref{corre})  into account, the same structure as $\psi(m)$ except that the spinors $u$ and $v$ are exchanged with the polarisation vectors $\e^r$ and except for the different allocation of factors of $s_i$. Here fermionic creation and annihilation operators are the $(\mathfrak{a}_i, \mathfrak{a}_i^\dagger)$ operators from (\ref{rolignu}), where one has to remember that the index '$i$' is a multi-index that also labels the generators of the Lie-group.  

The point here is that the commutator with the Bott-Dirac operator shifts between the bosonic and fermionic sectors, i.e. between the CAR algebra and the CCR algebra. 
The commutator with the Bott-Dirac operator will play an important role in section \ref{flucc}, where we consider inner fluctuations of the Bott-Dirac operator.

\section{Adding a Dirac operator on $M$}
\label{adding}

The algebraic representations related to quantum field theories, which we have discussed so far, all involve products $L^2(\Gamma)\otimes \Lambda^*(\Gamma)\otimes L^2(M)$ between a Hilbert space of states on a configuration space $\Gamma$ and a Hilbert space of functions on $M$. Since the Bott-Dirac operator acts only in the first Hilbert space it is natural to consider also a Dirac type operator, that acts in both Hilbert spaces. Let us therefore introduce a spinor bundle $S$ over $M$ and the Hilbert space $L^2(M,S)$ and write down the sum
\begin{equation}
D_{\mbox{\tiny\bf tot}} = B \otimes 1 + \gamma \otimes D
\label{extended}
\end{equation}
where $D$ is a Dirac operator acting on spinors in $L^2(M,S)$ and which depends on a metric on $M$ and where $\gamma$ is a suitable grading operator that satisfies $\{B,\gamma\}=0$ and $\gamma^2=1$.

Since $D$ has compact resolvent we can immediately conclude that $D_{\mbox{\tiny\bf tot}}$ has compact resolvent too. The question whether the commutator between $D_{\mbox{\tiny\bf tot}}$ and elements of the algebra ${\bf Alg}(\Gamma)$ is bounded or not. We suspect that this is not the case. The reason is the following: We can construct a state in $L^2 (\Gamma )$, which is peaked around a field configuration with fast oscillations. Hence for a suitably element in ${\bf Alg}(\Gamma)$ the corresponding function on $M$ generated by the representation given by the given field, will be  fast oscillating as well, and hence have a large commutator with    $D_{\mbox{\tiny\bf tot}}$. In particular the commutator will not be bounded. We thus suspect that 
$$
(D_{\mbox{\tiny\bf tot}}, {\bf Alg}(\Gamma), L^2(\Gamma)\otimes\Lambda^*\Gamma\otimes L^2(M,S))
$$
is not a semi-finite spectral triple.

\subsection{A conjectured link to the standard model}

Consider again the $\mathbf{HD}(M)$ algebra with the gauge group $SU(2)$ and let it be represented on spinors in $L^2(M,S)$, i.e. instead of the factor $\mathbb{C}^n$ in (\ref{ymm}) we use the spinor bundle $S$ for the representation. This means that the $\mathbf{HD}(M)$ algebra (as well as the $\mathbf{QHD}(M)$ algebra)  is represented in the Hilbert space
$$
\ch = L^2(\ca)\otimes \Lambda^*\ca \otimes L^2(M,S)
$$
where it now interacts with $D_{\mbox{\tiny\bf tot}}$ and where we by $\Lambda^*\ca$ again mean the inductive limit $\Lambda^*\mathbb{R}^n$. This geometrical construction is what we call 'quantum holonomy theory', which was first proposed in \cite{Aastrup:2015gba} and later developed in \cite{Aastrup:2017vrm,Aastrup:2017xde,Aastrup:2016caz,Aastrup:2016ytt}. 

The reason why we find this model particularly interesting is that it comes with a possible link to the formulation of the standard model of particle physics in terms of noncommutative geometry as developed by Chamseddine and Connes  \cite{Connes:1996gi,Chamseddine:2007hz}. The point is that if we restrict the algebra $\mathbf{HD}(M)$ to loops then it reduces in a classical limit characterised by a single (non-trivial) connection in $\ca$ to an almost commutative algebra
$$
\mathbf{HD}(M)\big\vert_{\mbox{\tiny loops}}\stackrel{\mbox{\tiny classical}}{\longrightarrow} C^\infty(M)\otimes M_2(\mathbb{C}).
$$
This correspondence, which was first pointed out in \cite{Aastrup:2012vq}, is is fact the main reason why we first became interested in holonomies, see \cite{Aastrup:2005yk}. Thus, in a semi-classical limit we have the general structure
$$
( D + D_F, C^\infty(M)\otimes M_2(\mathbb{C}), L^2(M,S)\otimes \ch_F )
$$
where $D$ is again the spatial Dirac operator and where $D_F$ is an operator that is given by the semi-classical limit of the Bott-Dirac operator and which will interact with the matrix factor $M_2(\mathbb{C})$. Also, by $\ch_F$ we refer to the Hilbert space $L^2(\ca)\otimes \Lambda^*\ca$ in the same limit. Thus, this argument suggest that something reminiscent of an almost-commutative spectral triple, which is the backbone of Chamseddine and Connes formulation of the standard model, will emerge from this model.

Clearly, more analyse is required in order to make this argument rigorous. Also, to fully compare this model with that of Chamseddine and Connes it would be useful to find a Hamiltonian formulation of the latter.


\section{Interactions and inner fluctuations}
\label{flucc}

Given a Dirac operator $D$ that interacts with an algebra $\cb$ it is natural to consider also the {\it fluctuated} Dirac operator \cite{marcolli}
$$
\tilde{D} = D + A, \quad A = \sum_i a_i [D, b_i], \quad a_i,b_i\in\cb
$$
where $A$ in the language of noncommutative geometry is a one-form. Now, in the present case we have the Bott-Dirac operator and an algebra ${\bf Alg}(\Gamma)$ of field operators (for example the algebra $C^\infty_b(M\times \mathbb{R})$ or the $\mathbf{HD}(M)$ algebra). We have already seen that the square of the Bott-Dirac operator gives rise to the free Hamiltonian in the field theories that we have considered. The question is, therefore, what the square of the {\it fluctuated} Bott-Dirac operator gives. As we shall see, the inclusion of the additional term gives rise to interactions, both bosonic and fermionic.

Let us begin with the Bott-Dirac operator and a representation of a general algebra ${\bf Alg}(\Gamma)$ of field operators where we have the field operator $\Theta$ 
and the commutator
$$
[B, \Theta] = \tilde{\Psi}
$$
where $ \tilde{\Psi}$ is a fermionic field as we discussed in section \ref{flu}. Thus, keeping the discussion at a general level we find the one-form
\begin{equation}
\Theta [B,  \Theta]  =  \Theta   \tilde{\Psi}
\label{plummer}
\end{equation}
and hence the {\it fluctuated} Bott-Dirac operator has the form
$$
\tilde{B} = B + \Theta   \tilde{\Psi}.
$$
Furthermore, its square gives
$$
\tilde{B}^2 = B^2 + H_{\mbox{\tiny fluc}}
$$
where the modification $H_{\mbox{\tiny fluc}}$ has the form
$$
H_{\mbox{\tiny fluc}} = (\Theta   \tilde{\Psi})^2 + \{B, \Theta   \tilde{\Psi}\}.
$$
Now, the point here is that while $B^2$ gives the Hamiltonians of the {\it free} system of bosonic and fermionic fields, then $H_{\mbox{\tiny fluc}} $ gives us interactions between the bosonic and fermionic sectors. This means that the construction with the Bott-Dirac operator is a general geometrical structure that based on an embedding of a configuration space $\Gamma$ produces well-defined interacting quantum field theories of bosonic and fermionic fields.


Let us also briefly discuss the more general case with the Dirac type operator $D_{\mbox{\tiny\bf tot}}$ as we discussed in section \ref{adding}. The one-form in (\ref{plummer}) then becomes\footnote{Note that this one-form is not expected to be bounded.}
$$
\Theta [D_{\mbox{\tiny\bf tot}},  \Theta]  =   \Theta   \tilde{\Psi}
 + \gamma \Theta [D, \Theta]
$$
leading to the {\it fluctuated} operator
$$
\tilde{D}_{\mbox{\tiny\bf tot}} = D_{\mbox{\tiny\bf tot}} +  \Theta   \tilde{\Psi}
 + \gamma \Theta [D, \Theta].
$$
Thus, if we consider the square 
$$
\tilde{D}^2_{\mbox{\tiny\bf tot}} = D^2 + B^2 + H'_{\mbox{\tiny fluc}}
$$
then we find a Hamilton operator that involves both a gravitational part (which remains classical), the square of the Bott-Dirac operator, that gives the Hamiltonian of the free bosonic and fermionic sectors, and finally the operator $H'_{\mbox{\tiny fluc}}$, which involves interactions that are both purely bosonic as well as mixed bosonic and fermionic. It is an interesting question precisely what kind of quantum field theories that will emerge from this geometrical setup, both in the case of a scalar theory and in the case of a gauge theory.

\section{Locality and the question of quantum gravity}
\label{locality}

In the quantum field theories, that we have discussed so far, the expectation values in $L^2(\Gamma)$, where $\Gamma$ is a configuration space, all have the general form
$$
\langle \eta(\psi)\vert \OO(\psi)\vert \eta(\psi)\rangle_{L^2(\Gamma)} = \int_{\Gamma} d\psi \OO(\psi) \exp( - \Vert \oo_p\psi\Vert_{\mbox{\tiny s}} )
$$
where $\OO(\psi)$ is some composite field operator.
The point here is that the Sobolev norm $\Vert\cdot\Vert_{\mbox{\tiny s}}$ plays the role of a weight in the measure 
over $\Gamma$. This implies that field configurations, that have a small Sobolev norm, will have a larger weight compared to field configurations, that have a large Sobolev norm. Put differently, field configurations, that vary at a large scale have a larger weight than those that vary at a short scale, and in particular it means that field configurations, that are singular, have zero measure as such configurations will have infinite Sobolev norm\footnote{This is best seen by the fact that the Sobolev norm, that is relevant for our case, dominates the supremum norm in three dimensions \cite{Nirenberg}.}. Thus, extreme situations, such as the initial big bang singularity and big bang singularities appear to be ruled out within this construction.

This means that the quantum field theories, which we have presented, come with an ultra-violet suppression that is enforced by representation theory. This raises an interesting question concerning the search for a theory of quantum gravity.

It is generally believed that a theory of quantum gravity will in one way or another quantise space and time and that quantum effects of the gravitational fields will play an important role below the Planck length. Simple arguments, that combine quantum mechanics and general relativity, suggest that measurements below the Planck length are operational meaningless since the measuring probe must carry so much energy and momentum that it will create a black hole, and a theory of quantum gravity is believed to be the source of such an short scale fix. But if quantum field theory effectively dampens degrees of freedom below the Planck length -- as it happens within the framework, that we have presented -- then one might ask whether the gravitational field needs to be quantised at all? If quantum field theory suppresses those very field configurations, that would otherwise probe the quantum domain of the gravitational field, then it seems to us that one of the key arguments in favour of a theory of quantum gravity would be invalidated.\\

\section{Discussion}

One of the most interesting feature of the quantum field theories, that we have presented in this paper, is that they exist in a rigorous sense. 
To the best of our knowledge these are the first examples of interacting quantum field theories in $3+1$ dimensions to have this feature. 
What we find is that quantum field theory can be understood as a geometrical construction over the appropriate space of field configurations. In the cases that we analyse --  a real scalar field and a gauge field, both with and without a fermionic sector -- we find that the free sectors of these theories are generated by a universal linear Bott-Dirac operator, that intertwines the CAR and the CCR algebras and interacts with an appropriate algebra of field operators. The interacting theories are then perturbations of this geometrical construction.

It is clear that although one can use this framework to engineer probably any variety of quantum field theory there exist  a particular class of theories, which are singled out by the construction of the Bott-Dirac operator. The key question is what algebra ${\bf Alg}(\Gamma)$ of field operators one should choose. Once this choice has been made everything appears to follow essentially canonically. We have argued that the $\mathbf{HD}(M)$ algebra is a particularly natural choice but it may be that other choices will prove even more interesting. Here the ultimate criteria of success must be whether a particular choice of algebra is able to connect to and explain the structure of the standard model of particle physics.

What remains now is to understand precisely what these quantum field theories contain and how they incorporate the known features of ordinary perturbative quantum field theory and where they differ.

Here a key question is that of space-time covariance. A (perturbative) quantum field theory on a curved background is usually formulated with respect to a space-time background metric where the causal structure of the metric is build into the quantum theory via the commutation relations. In the present case we start out with a 3-dimensional manifold with a metric. The question is, therefore, whether the quantum theory, that we find, will be covariant with respect to the 4-metric that emerges with time evolution. It is interesting that this metric appear to automatically have the Minkowski signature, which suggest that special relativity could be an output of this framework. 

Another question is whether the non-locality, that characterises the representations, which we find, will affect the causal structure of the emergent space-time manifold, i.e. whether the non-locality will be accompanied with a correction to special relativity.

The representations of algebras of field operators, that we find, all depend on a scale $\tau_1$ -- we interpret it as the Planck scale -- where degrees of freedom are dampened according to their relation to this scale and where field operators can only be localised up to corrections proportional to $\tau_1$. This means that these representations cease to exist in the local limit $\tau_1\rightarrow 0$, which is of course the limit where the UV divergencies known from perturbative quantum field theory arise. This raises the question how renormalisation theory will emerge in this limit and what role it plays away from $\tau_1=0$. 

A related question is that of gauge invariance. The measure, which we construct for the ${\mathbf{HD}(M)}$ algebra, is not gauge invariant as different elements in a gauge orbit will have different Sobolev norms. We think this can be understood in terms of a gauge fixing procedure but more analysis is required to fully understand this issue.


As already mentioned the ultimate test for any candidate for a fundamental theory must be whether it can explain the structure of the standard model of particle physics. Here we believe that the $\mathbf{HD}(M)$ algebra is particularly interesting as it produces an almost commutative algebra in a semi-classical limit, which is the type of algebra that Chamseddine and Connes have identified as the cornerstone in their formulation of the standard model  \cite{Connes:1996gi,Chamseddine:2007hz}. Again, to pursue this line of analysis we will need to employ the full toolbox of noncommutative geometry as well as carefully analysing the semi-classical limit. In respect to the latter it is interesting that Higson and Kasparov \cite{Higson} have already developed such an analysis, albeit for a different purpose.

Note that both the CCR and the CAR algebra appear naturally in our construction. Due to the construction of $L^2(\Gamma )$ we have already chosen a representation of the CCR algebra. As for the CAR algebra we have in this article notationally chosen the Fock space as the representation but we are free to choose any representation we would like. The CAR algebra has representations, which in the weak closure give type III von Neumann algebras. This  opens up for applying Tomita-Takesaki theory, where we would get a one-parameter groups of automorphisms, which would entail a time development. It is interesting to compare this to the dynamics given by the Hamiltonian.

It is striking that the quantum field theories, which we find, appear to solve the problem of space-time singularities, a problem that is normally expected to be solved by a theory of quantum gravity. If states cannot be arbitrarily localised in space it means that those field configurations, which would cause space to curve infinitely, cannot form. This seems on the one hand to prevent the singularities purported to exist within black holes and at the initial big bang and on the other hand to remove one of the most important arguments in favour of a theory of quantum gravity. We find this idea rather interesting.

Let us end this paper on a speculative note. One of the most interesting problems in contemporary theoretical physics is the question of the mass gap in Yang-Mills theories. Since the square of the Bott-Dirac operator, which gives the free Hamiltonian for a non-perturbative Yang-Mills theory, has a discrete spectrum with the lowest eigenvalue being zero, one might think that this holds also for a full, interacting theory. If this is the case then we believe that we will have the interesting situation that the mass gap contains information about the size of the universe -- in the sense that the spectrum of the square of the Bott-Dirac operator becomes continuous in the limit where the manifold is no longer compact. Put differently, the existence of the mass gap could be read as evidence that our spatial universe is compact.

\vspace{1cm}
\noindent{\bf\large Acknowledgements}\\

We would like to thank Prof. Nigel Higson for bringing his work with Prof. Gennadi Kasparov on the Bott-Dirac operator to our attention.
JMG would like to express his gratitude towards Ilyas Khan, United Kingdom, for his generous financial support. JMG would also like to express his gratitude towards the following sponsors:  Ria Blanken, Niels Peter Dahl, Simon Kitson, Rita and Hans-J\o rgen Mogensen, Tero Pulkkinen and Christopher Skak for their financial support, as well as all the backers of the 2016 Indiegogo crowdfunding campaign, that has enabled this work. Finally, JMG would like to thank the mathematical Institute at the Leibniz University in Hannover for kind hospitality during numerous visits.\\


\begin{thebibliography}{99}







\bibitem{Aastrup:2017vrm}
  J.~Aastrup and J.~M.~Grimstrup,
  ``Representations of the Quantum Holonomy-Diffeomorphism Algebra,''
  arXiv:1709.02943.
  
  
\bibitem{Aastrup:2017xde}
  J.~Aastrup and J.~M.~Grimstrup,
  ``Quantum Gravity and the Emergence of Matter,''
  arXiv:1709.02941.

  
  
\bibitem{Higson}
N.~Higson and G.~Kasparov, "E-theory and KK-theory for groups which act properly and isometrically on Hilbert space", 
Inventiones Mathematicae, vol. {\bf 144}, issue 1, pp. 23-74.




\bibitem{AGnew}
  J.~Aastrup and J.~M.~Grimstrup,
  ``C*-algebras of Holonomy-Diffeomorphisms and Quantum Gravity II'',
   J.\ Geom.\ Phys.\  {\bf 99} (2016) 10.






\bibitem{Aastrup:2015gba}
  J.~Aastrup and J.~M.~Grimstrup,
  ``Quantum Holonomy Theory,''
  Fortsch.\ Phys.\  {\bf 64} (2016) no.10,  783.





\bibitem{Connes:1996gi}
  A.~Connes,
  ``Gravity coupled with matter and foundation of noncommutative geometry,''
  Commun.\ Math.\ Phys.\  {\bf 182} (1996) 155.



\bibitem{Chamseddine:2007hz}
  A.~H.~Chamseddine and A.~Connes,
  ``Why the Standard Model,''
  J.\ Geom.\ Phys.\  {\bf 58} (2008) 38.






\bibitem{Aastrup:2012vq}
  J.~Aastrup and J.~M.~Grimstrup,
  ``C*-algebras of Holonomy-Diffeomorphisms and Quantum Gravity I,''
  Class.\ Quant.\ Grav.\  {\bf 30} (2013) 085016.





\bibitem{Aastrup:2005yk}
  J.~Aastrup and J.~M.~Grimstrup,
  ``Spectral triples of holonomy loops,''
  Commun.\ Math.\ Phys.\  {\bf 264} (2006) 657.








\bibitem{Aastrup:2012jj}
  J.~Aastrup and J.~M.~Grimstrup,
  ``Intersecting Quantum Gravity with Noncommutative Geometry: A Review,''
  SIGMA {\bf 8} (2012) 018.



 
\bibitem{Aastrup:2010ds}
  J.~Aastrup, J.~M.~Grimstrup and M.~Paschke,
  ``Quantum Gravity coupled to Matter via Noncommutative Geometry,''
  Class.\ Quant.\ Grav.\  {\bf 28} (2011) 075014.

  
  

\bibitem{Aastrup:2009et}
  J.~Aastrup, J.~M.~Grimstrup, M.~Paschke and R.~Nest,
  ``On Semi-Classical States of Quantum Gravity and Noncommutative Geometry,''
  Commun.\ Math.\ Phys.\  {\bf 302} (2011) 675.





\bibitem{Aastrup:2009ux}
  J.~Aastrup, J.~M.~Grimstrup and R.~Nest,
  ``Holonomy Loops, Spectral Triples \& Quantum Gravity,''
  Class.\ Quant.\ Grav.\  {\bf 26} (2009) 165001.




\bibitem{Aastrup:2008zk}
  J.~Aastrup, J.~M.~Grimstrup and R.~Nest,
  ``A New spectral triple over a space of connections,''
  Commun.\ Math.\ Phys.\  {\bf 290} (2009) 389.




\bibitem{Aastrup:2008wa}
  J.~Aastrup, J.~M.~Grimstrup and R.~Nest,
  ``On Spectral Triples in Quantum Gravity I,''
  Class.\ Quant.\ Grav.\  {\bf 26} (2009) 065011.




\bibitem{Aastrup:2008wb}
  J.~Aastrup, J.~M.~Grimstrup and R.~Nest,
  ``On Spectral Triples in Quantum Gravity II,''
  J.\ Noncommut.\ Geom.\  {\bf 3} (2009) 47.
















\bibitem{Aastrup:2016caz}
  J.~Aastrup and J.~M.~Grimstrup,
  ``Quantum Holonomy Theory and Hilbert Space Representations,''
  Fortsch.\ Phys.\  {\bf 64} (2016) 903.


\bibitem{Aastrup:2016ytt}
  J.~Aastrup and J.~M.~Grimstrup,
  ``On a Lattice-Independent Formulation of Quantum Holonomy Theory,''
  Class.\ Quant.\ Grav.\  {\bf 33} (2016) no.21,  215002.






\bibitem{Aastrup:2014ppa}
  J.~Aastrup and J.~M.~Grimstrup,
  ``The quantum holonomy-diffeomorphism algebra and quantum gravity,''
  Int.\ J.\ Mod.\ Phys.\ A {\bf 31} (2016) no.10,  1650048.


  
  


  

 




\bibitem{Nirenberg}
L.~Nirenberg, "On elliptic partial differential equations." Annali della Scuola Normale Superiore di Pisa\ -\ Classe di Scienze {\bf 13.2} (1959): 115-162. 




\bibitem{Campagnari:2009km}
  D.~R.~Campagnari, H.~Reinhardt and A.~Weber,
  ``Perturbation theory in the Hamiltonian approach to Yang-Mills theory in Coulomb gauge,''
  Phys.\ Rev.\ D {\bf 80} (2009) 025005.
  

























\bibitem{marcolli}
  A.~Connes and M.~Marcolli,
  ``Noncommutative Geometry, Quantum Fields and Motives,''
  Colloquium Publications, 
Volume: {\bf 55} (2008).
Print ISBN: 978-0-8218-4210-2.
  




\end{thebibliography}
\end{document}